\documentclass{article}
\date{}
\usepackage{float}
\vspace{1cm}
\usepackage[english]{babel}

\usepackage[a4paper,top=2cm,bottom=2cm,left=3cm,right=3cm,marginparwidth=1.75cm]{geometry}

\usepackage{amsmath}
\usepackage{graphicx}
\usepackage[colorlinks=true, allcolors=blue]{hyperref}

\title{\textbf{A new method of petroleum well logging
— Electrical impedance logging}}
\author{Weinan Wang}

\begin{document}
\maketitle

\begin{abstract}
  This paper presents a new petroleum well logging method—electrical impedance logging for shaly sand reservoirs—through theoretical and petrophysical experimental research. Electrical impedance logging measures the electrical impedance of shaly sand reservoirs, extracts resistivity information from the real part, and uses it to determine the oil saturation of the reservoir quantitatively. The study shows that the resistivity, extracted from the real part of the electrical impedance in shaly sands, has characteristics similar to those of the pure sandstone formation resistivity and can be directly used in Archie's law for oil-bearing interpretation of reservoirs.

\end{abstract}

\section{Introduction}

\hspace{\parindent}Resistivity logging is one of the most widely used well logging methods. Archie established a quantitative relationship between reservoir resistivity and oil saturation[1], which remains the best method for evaluating the oil-bearing properties of reservoirs to this day. Archie's law was established under conditions of relatively uniform lithology, absence of clay, and well-developed porosity. Over the past decades, logging analysts have been striving to establish models for oil saturation in shaly sand reservoirs, trying various methods and publishing numerous interpretation methods and models, such as the Waxman-Smits model[2] and the Dual-Water model[3]. However, none of these methods or models have universal applicability, and extensive research is still ongoing. The difficulty lies in the fact that most of these interpretation methods or models are based on regional geology or are empirical, and thus lack universality. Additionally, some parameters in these models are difficult to determine. Therefore, the electrical impedance well logging method proposed in this paper can better address the aforementioned issues.

The electrical impedance well logging method can provide the X-component information of the formation at low frequencies, with the measurement frequency selected at 1 kHz, which is consistent with the applicable frequency of Archie's law. Among the existing logging methods, the coil method, such as electromagnetic wave logging and induction logging, is one of the few methods that can provide the X-component information of the formation. The coil method instruments can achieve higher accuracy only at higher frequencies. For example, the frequency of electromagnetic wave logging is generally above tens of MHz, which is mainly used to provide the dielectric constant of the formation.The frequency of induction logging is generally 20 kHz, in most cases, the resistivity obtained cannot be directly used in Archie's equation but needs to be corrected for frequency dispersion. In recent years, some oil fields in China have conducted complex resistivity logging, but this method cannot provide the X-component information of the formation.

\section{The proposal and theoretical foundation of the electrical impedance well logging methodles}

\hspace{\parindent}Electrical impedance well logging is proposed for the evaluation of oil saturation in shaly sand reservoirs. By measuring the electrical impedance of the reservoir, the real part resistivity of the impedance is extracted, which can be regarded as the resistivity of pure sandstone formations under the same conditions. That is, the real part resistivity extracted from the reservoir impedance is considered to represent the resistivity of a pure sandstone formation with the same porosity, permeability, pore structure, connectivity, etc.This allows for the direct quantitative evaluation of the oil saturation in the reservoir using Archie's law. Here, clay is considered as part of the rock matrix, and the contribution of the clay to the rock resistivity is fully reflected in the imaginary part of the impedance.

At low frequencies, the admittance of the rock can be expressed in the following complex form.
 
\[ \textbf{\textit{Y=C+jB}\textit{}} \quad\quad\quad (1)\]    
\hspace{\parindent}In formula (1), \textit{\textbf{Y}} represents the admittance of the rock, \textbf{\textit{C}} is the real part of the rock admittance, equivalent to the conductance of a pure sandstone reservoir with the same porosity, permeability, pore structure, and connectivity, and it represents the contribution of ionic conduction current; \textbf{\textit{B}} is the imaginary part of the rock admittance, mainly representing the contribution of displacement current, which varies with the clay content of the formation and the measurement frequency.

In the logging interpretation of shaly sands, the increased conductivity caused by clay is referred to as additional conductance; see formula (2).      

\[ \textbf{{\textbf{\textit{C}}r=\textbf{\textit{C}}+\textbf{\textit{A}}}\textbf{}} \quad\quad (2)\]
\par
In formula (2), \textbf{\textit{Cr}} is the conductance of the reservoir rock, \textbf{\textit{C }}is the same as in formula (1), which refers to the conductance of pure sandstone formation under the same conditions, and \textbf{\textit{A}} represents the additional conductance caused by clay.
\par
From equations (1) and (2), we can derive equations (3) and (4).

\[ \textit{C}\textbf{=\textit{\textbf{Y}} - }\textit{jB}  \quad\quad \quad (3)\]
\[ \textbf{\textit{C=Cr - A}}  \quad\quad \quad       (4)\]
\par
Equations (3) and (4) have the same form, with \textbf{\textit{C}} in both equations representing the conductance of pure sandstone that can be directly used in Archie's equation.
\par
As can be seen, the parameters in equation (3) can be obtained by measuring the admittance of the rock, with the imaginary part being related to clay content and measurement frequency. Any changes in the rock's conductance due to the presence of clay are reflected in the imaginary part. However, \textbf{\textit{A}} in equation (4) is a parameter derived from experimental analysis, statistical methods, or empirical approaches, and it is calculated differently across various regions and interpretative models; therefore, the results are also different. The model based on equation (4) cannot universally reflect changes in reservoir conductance caused by variations in clay content. Additionally, some parameters in certain models are difficult to determine.
\par
The physical quantity used in well logging interpretation is rock resistivity. By transforming equation (1) into impedance form, refer to formula (5).

\[ \textbf{\textit{Z=R+jX}}  \quad\quad \quad           (5)\]

\par
In the formula (5), \textbf{\textit{Z}} is the electrical impedance of the formation, \textbf{\textit{R }}is the real part of the rock impedance (\textbf{\textit{R}}-component), which is equivalent to the resistance of a pure sandstone reservoir with the same porosity, permeability, pore structure, and connectivity, it represents the contribution of ionic conduction current; \textbf{\textit{X}} is the imaginary part of the rock impedance (\textbf{\textit{X}}-component), mainly representing the contribution of displacement current, which varies with the clay content of the formation and the measurement frequency.
\par
In homogeneous and pure reservoirs with low clay content, the \textit{\textbf{X}} signal is weaker, and the reservoir resistivity obtained from the \textit{\textbf{R}}-component is not significantly different from that measured by conventional methods. However, in shaly sand reservoirs with higher clay content, the impact of the \textbf{\textit{X}} signal cannot be ignored because of the presence of clay, leading to a significant difference between the formation resistivity measured by conventional methods and that obtained from the \textbf{\textit{R}}- component in equation (5). Under these conditions, the resistivity extracted from the \textit{\textbf{R}}-component should be used for the well logging interpretation of reservoir oil saturation, which yields better results.
\par
The formation resistivity \(\textbf{\textit{R}}_\textbf{\textit{t}}\) can be obtained by multiplying {\textbf{\textit{R}}} in equation (5) by the electrode coefficient, as shown in formula (6). This resistivity differs from the formation resistivity measured by conventional methods using the same electrode system and frequency. It eliminates the effect of clay on the formation resistivity, which is reflected in the imaginary part.
\par
\[\textbf{\textit{R}}_t=\textbf{\textit{K}}\times\textbf{\textit{R}}\quad\quad \quad(6)\]

\par
In formula (6),\(\textbf{\textit{R}}_\textbf{\textit{t}}\) is the resistivity of the reservoir, \textbf{\textit{R}} is the resistance of the reservoir, and \textbf{\textit{K }}is the well logging electrode coefficient.
\par
\

Therefore, the oil saturation of the reservoir can be directly calculated using Archie's equation, as shown in formula (7).

\[\textbf{\textit{S}}^{n}_{w}=\dfrac{ab\textbf{\textit{R}}_w}{\phi^m\textbf{\textit{R}}_t} \quad\quad \quad(7)\]

\vspace{2mm}
\par

In formula (7), $\textbf{\textit{S}}_w$ is the water saturation of the reservoir, $\textbf{\textit{R}}_t$ is the real part of the resistivity extracted from the reservoir electrical impedance, $\textbf{\textit{R}}_w$ is the formation water resistivity, \(\phi\) is the effective porosity of the reservoir, and a, b, m, and n are coefficients related to the reservoir lithology, which generally need to be determined by petrophysical experiments.

In the shaly sands reservoirs, the surfaces of the clay particles are negatively charged. According to electrochemical theory, the negative charges on the surfaces of the clay particles adsorb cations from the pore water, forming a double electrical layer. The capacitive property of the shaly sands reservoirs is the macro manifestation of these countless microscopic double layers. Under the influence of an electric field, the negatively charged surfaces of the clay particles do not move, but the adsorbed cations on their surfaces move towards the positive electrode, forming a current. The conduction mechanism is similar to that of hole conduction in semiconductor materials. According to the law of independent ion movement, the conduction current produced by the migration of ions in the pores and the displacement current produced by the clay are relatively independent contributors to the electrical conductivity of the rock. If the clay is regarded as part of the rock matrix, the relationship between the conduction current of the pore fluid and the effective porosity follows Archie's law.
\par
Some literature supports the views of this paper. M. H. Waxman and H. J. Vinegar believe that the same-phase saturation index of shaly sands is similar to Archie's saturation index for pure sandstone[4].
\par
In their paper on the dual water model, C. Clavier, G. Coates, and J. Dumanoir believe that from the perspective of conductivity, shaly formations are characterized by a mixture of formation water and clay-bound water, similar to pure formations[3]. In terms of petrophysical properties, shaly formations can be characterized by total porosity, formation factor, water saturation, total conductivity, and the equilibrium ion concentration of clay per unit pore volume; these characteristics are somewhat similar to those of pure formations with the same parameters. They also believe that the close relationship between the formation factor of the dual water model and Archie's formation factor is significant in log interpretation.
\par
Juhasz also believes that shaly formations are generally evaluated based on the total porosity and effective porosity system, and the difference between the total porosity and effective porosity is the clay-bound water volume[5], as shown in Fig. 1. This view is consistent with the dual water model.

\begin{figure}[htbp]   
            \centering
            \includegraphics[width=0.5\linewidth]{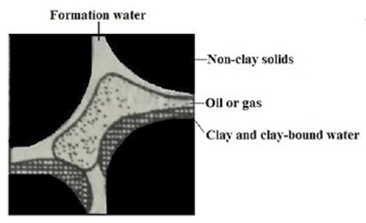}
                \caption{ Schematic diagram of rock pore structure}
    \label{fig:enter-label}
\end{figure}
\par
The viewpoints of the above references are consistent with this paper, as they all recognized that conduction current should be distinguished from displacement current. However, their methods differ from those in this paper. They use experimental analysis, statistical methods, or empirical approaches to establish interpretative models, attributing the effects of displacement current to these models. Due to the limitations of regional geological conditions, the interpretation model of one area is difficult to apply to another, and some models include many theoretical or experimental parameters that are challenging to determine. In contrast, this paper measures formation electrical impedance, extracts resistivity information from its real part, and then directly applies Archie's law to quantitatively determine the oil saturation of the reservoir, thereby compensating for the limitations of the aforementioned methods.
\par
There are also references that believe that in medium and low resistivity formations greater than 1.5 m, the resistivity measured by induction logging can be regarded as true resistivity, and Archie's law can be used directly for oil-bearing evaluation without any correction[6], which is consistent with the view of this paper.
\par

\section{Experimental verification of the electrical impedance 
well logging method}
 
 \par
 
\hspace{\parindent}Figure 2 shows the relationship between the electrical impedance at 1 kHz and the electrical impedance at 20 kHz (the left figure represents the real part, while the right figure represents the imaginary part). It can be seen that there is a good linear relationship between the real parts of the impedance at the two frequencies; however, the correlation between the imaginary parts is weaker, indicating that the effect of clay on the impedance of rocks is primarily reflected in the imaginary part.
                                   
\par

\begin{figure}[htbp]    

                                   \centering                                 \includegraphics[width=0.75\linewidth]{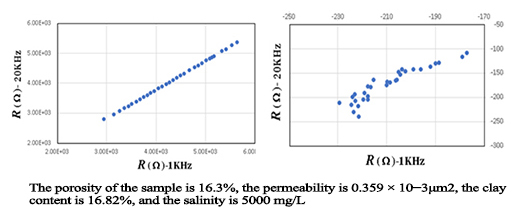}
                                                   
                                     \caption{The relationship between the electrical impedance at 1 kHz and at 20 kHz}   
    \label{fig:enter-label}
\end{figure}
       
  \par                   
Fig. 2 also shows that for the induction log (20 kHz), the measured resistivity in medium and low resistivity formations can be regarded as true resistivity, and Archie's law can be used directly for oil-bearing evaluation without any correction, which confirms the conclusion of reference 6 and is consistent with the view of this paper.
\par
 Fig.3 shows the relationship between clay distribution and its effect on the dielectric constant and resistivity of artificial rock samples at a measurement frequency of 25 MHz[7]. The artificial samples were created by mixing fine sand and clay in various proportions. It can be observed that clay distribution significantly impacts the dielectric constant of the samples but has minimal influence on resistivity. This result indicates that clay distribution has a smaller effect on the real part of electrical impedance and a larger effect on the imaginary part, further suggesting that the impact of clay on the rock's electrical impedance is primarily reflected in the imaginary part. Therefore, it is feasible to equate the real part of the electrical impedance to the resistance of a pure sandstone formation.

\begin{figure}[htbp]                            
                     \centering
                        \includegraphics[width=0.75\linewidth]{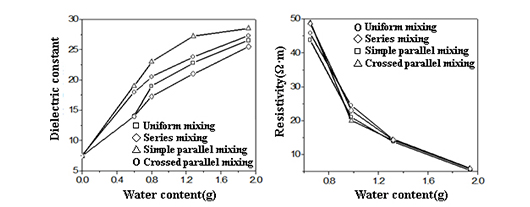}
                      
                        \caption{Influence of clay distribution in artificial samples on their dielectric constant and resistivity}
      
                \label{fig:enter-label}
    \end{figure}
        
\newpage

Table 1 compares the resistivity obtained from the Waxman-Smits model (W-S model) for the same rock sample with the resistivity extracted from the real part of the impedance, showing an average difference of 11.21\%. The W-S model is considered a good model for evaluating the conductivity of shaly sand reservoirs, and the comparison results in Table 1 indicate a comparatively good correlation between the resistivity derived from the W-S model and the resistivity extracted from the real part of the impedance.Although this comparison is not strictly accurate due to the lack of pure sandstone samples as a standard, it demonstrates the validity of the research direction.
\par

 \begin{tabular}[h]{cccc}
 \\
        \hline
        \centering
        Sample & Porosity (\%) & W-S model $R_o$ ($\Omega\cdot$m) & Impedance $R_o$ ($\Omega\cdot$m) \\
        \hline
        1  & 23.1 & 13.66 & 11.58 \\
        2  & 23.8 & 12.90 & 11.63 \\
        3  & 22.4 & 10.73 & 11.95 \\
        4  & 21.6 & 12.77 & 16.43 \\
        5  & 25.1 & 10.63 & 10.54 \\
        6  & 25.6 & 11.04 & 13.71 \\
        7  & 16.9 & 22.29 & 20.76 \\
        8  & 18.8 & 17.66 & 15.56 \\
        9  & 24.1 & 12.49 & 12.08 \\
        10 & 19.9 & 15.59 & 14.06 \\
        11 & 22.4 & 11.66 & 11.53 \\
        12 & 23.0 & 14.28 & 12.31 \\
        \hline
        \\
\end{tabular}
 
\par

 Tab.1 Comparison of resistivity obtained from the W-S model with those extracted by the impedance method
\par
\section{Design of the electric impedance logging electrode system}
\par
\hspace{\parindent}This paper employs the Wheatstone bridge method to measure the electrical impedance of formations. The Wheatstone bridge method boasts high accuracy in measuring electrical impedance and is commonly utilized in laboratories for measuring the electrical impedance of rock samples. This technology is already relatively mature.
\par
Fig.4 shows the schematic diagram of the electrical impedance well logging electrode system.
\par

The measurement principle is roughly as follows: the target formation is regarded as the measured arm \textbf{\textit{Zx}} of the automatic balancing bridge, which is connected to the measurement circuit. By automatically adjusting the impedance of the unknown bridge arm \textbf{\textit{Zs}}, the current in the galvanometer\textit{\textbf{ G }}is made zero, and the bridge is balanced. The real and imaginary parts of the \textbf{\textit{Zs}} impedance correspond to the real part (resistance) and imaginary part (reactance) of the formation impedance.
\par

\begin{figure}[htbp]
                          \centering
                            \includegraphics[width=0.5\linewidth]{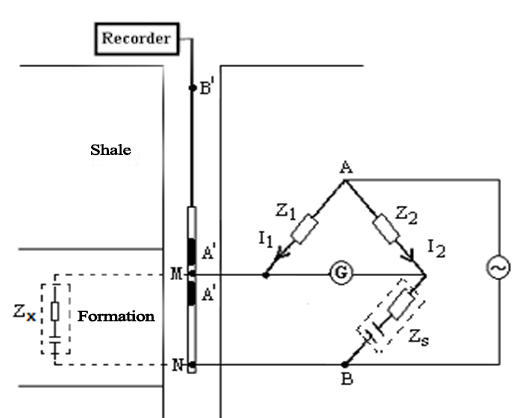}               
                \caption{Schematic diagram of the electrical impedance well logging electrode system}(A and B are power supply electrodes, M and N are measurement electrodes, A' is the focusing electrode, and B' is the return electrode of the focusing loop)
            \label{fig:enter-label}
        \end{figure}

The reservoir resistivity can be obtained from Equation (8).
\par

\[R_t=\textbf{\textit{K }}\dfrac{\textbf{\textit{U}}}{\textbf{\textit{I}}_1}   \quad\quad \quad    (8)\]
\par

Where \(\textit{\textbf{R}}_\textit{t}\) is the formation resistivity value extracted from the real part of the electrical impedance, and is regarded as the resistivity of a pure sandstone reservoir with the same porosity, permeability, pore structure, connectivity, etc.\(\textbf{\textit{U}}\) is the voltage applied across the two ends of the bridge's unknown arm (the voltage between M and N). \(\textbf{\textit{I}}_1\) is the current flowing through the bridge's unknown arm (the formation), and \textbf{\textit{K}} is the electrode coefficient.
\par
After preliminary analysis of the detection characteristics of the electrode system, when the distance between electrodes A and B exceeds 6 meters, both the radial detection depth and vertical resolution of the instrument can reach 0.5 meters.
\par
It can be seen that the electric field distribution, resistivity calculation, electrode coefficient determination, influencing factors, and correction methods of electrical impedance well logging are all consistent with those of the laterolog, but the measurement loop also includes an imaginary part measurement function. The laterolog is a mature technology, and laboratory measurements of core sample electrical impedance are also well-established. Therefore, it is believed that this electrode system can be used to measure the electrical impedance of the formation.
\par
When the laboratory measures the electrical impedance of the sample, the current passes through the core (inside), while in field logging, the instrument moves within the wellbore to measure (outside). Applying laboratory measurement technology to field logging requires significant effort; therefore, this design is preliminary, and specific issues need to be addressed continuously.
\par

\section{Conclusion}
\par
\hspace{\parindent}The electrical impedance logging proposed in this paper is a new well logging method for shaly sand reservoirs. Research shows that resistivity extracted from the real part of the shaly sand electrical impedance has characteristics similar to those of pure sandstone formation resistivity and can be used directly with Archie's law for reservoir oil content interpretation. The well logging electrode system designed using the Wheatstone bridge method can meet the requirements for measuring formation electrical impedance. 

\par
\section{\textbf{Acknowledgements}}
\hspace{\parindent}Thanks to Zhang Yu Jin from 3G Land (Beijing) Technology Pty. Ltd. for helping to process the relevant materials.

\par
\section{References}
\par

\hspace{\parindent}[1].Archie G E.The eLectrlcal resistivity log as an aid in determining some reservoir characteristics. AIME,146(1):5462,1942.
\par
[2]Waxman M.H. and Smith L.J.M.Electrical Conductivities in Oil-bearing Shaly Sands. Society of Petroleum Engineers Journal, June, p107-122,1968.
\par
[3]C.Clavier,G.Coates,J.Dumanoir.The theoretical and experimental bases for the "dual water" model for the interpretation of shaly sands.Soc. Pet. Eng. J.,(Apr.): 153-168,1984.
\par
[4]M.H.Waxman and H.J.Vinegar.Complex conductivity model of shaly sands. Translated by Wang Wenxiang. Evaluation of shaly sands, Information Research Institute of China National Petroleum Corp:p76-84,1989.
\par
[5]Juhasz. The Central Role Of Qv And Formation-water Salinity In The Evaluation Of Shaly Formations.SPWLA TWENTIETH ANNUAL LOGGING SYMPOSIUM, June, 3-6.1979.
\par
[6]Chen Yiming. Well Logging Methods and Data Applications, Jianghan Petroleum College Textbook, 1992.
\par
[7]Wang weinan, Zhang xi, et.al. Experimental Investigation of Dielectric Constant for Low Porosity and Permeability. SPE 30871,1997.

\end{document}